\RequirePackage[2020-02-02]{latexrelease}
\documentclass{iau_main}

\usepackage{amsmath}
\usepackage{graphicx}
\usepackage{multirow}
\usepackage{multicol}

\begin{document}

\lefttitle{Janhavi Baghel et al.}
\righttitle{Radio Polarization: A Powerful Resource for Understanding the Blazar Divide}

\jnlPage{1}{7}
\jnlDoiYr{2021}
\doival{10.1017/xxxxx}

\aopheadtitle{Proceedings IAU Symposium}
\editors{eds.}

\title{Radio Polarization: A Powerful Resource for Understanding the Blazar Divide}

\author{Janhavi Baghel$^{1}$,
P. Kharb,$^{1}$
Silpa S.,$^{1}$
Luis C. Ho,$^{2, 3}$
C. M. Harrison$^{4}$}
\affiliation{
$^{1}$National Centre for Radio Astrophysics (NCRA) - Tata Institute of Fundamental Research (TIFR), S. P. Pune University Campus, Ganeshkhind, Pune 411007, India \\
 $^{2}$Kavli Institute for Astronomy and Astrophysics, Peking University, Beijing 100871, China\\
 $^{3}$Department of Astronomy, School of Physics, Peking University, Beijing 100871, China\\
 $^{4}$School of Mathematics, Statistics and Physics, Newcastle University, Newcastle upon Tyne NE1 7RU, UK}

\begin{abstract}
With high-sensitivity kiloparsec-scale radio polarimetry, we can examine the jet-medium interactions and get a better understanding of the blazar divide in radio-loud (RL) AGN. We are analyzing the radio polarimetric observations with the EVLA and GMRT of 24 quasars and BL~Lacs belonging to the Palomar-Green (PG) sample. The RL quasars show extensive polarisation structures in their cores, jets, lobes, and hotspots, whereas preliminary results suggest that BL~Lacs exhibit polarisation primarily in their cores and inner jet regions. These findings imply that both intrinsic (central engine-related) and extrinsic (environment-related) variables are important in the formation of the blazar subclasses. The Fanaroff-Riley (FR) dichotomy can also be studied assuming RL unification and looking through the lens of blazars. Due to the radio-unbiased nature of the optically/UV-selected PG sample, we find a large fraction of the PG quasars are restarted, distorted (S- or X-shaped), or have a hybrid FR morphology.
\end{abstract}


\maketitle
\section{Introduction}
Active galactic nuclei (AGN) are supermassive black holes \citep[SMBH;][]{Rees84} actively accreting matter to form the highly energetic, bright, non-stellar centres of galaxies. Relativistic jets of synchrotron-emitting plasma are sometimes launched orthogonal to the accretion disks \citep{Blandford1974}. About $10-15\%$ of AGN are radio-loud (RL) AGN parameterised as having {${R = (S_{5~GHz}/S_{{B}-band}) \geq 10}$} where $R$ is the ratio of radio ($5~GHz$) to optical ({\it B}-band) flux densities \citep{Kellermann1989}. They have large-scale jets which can extend up to megaparsec scales and significantly affect their host galaxy and surroundings, influencing star formation, heating up the circumgalactic medium, and galaxy evolution \citep[][]{Blandford2019}. Differences in observational properties have led to an intrinsically anisotropic model of the physical structure of AGN. The orientation-based RL unification hypothesis \citep{Urry95} suggests that these differences can be explained by changes in the angle at which an observer sees the AGN, obscuration by an opaque dust ring \citep[called torus;][]{Antonucci1985} and the impact of relativistic beaming. Different classes of RL AGN, viz., radio galaxies (RGs), quasars and BL Lac objects, distinguished primarily by their optical emission lines, are thought to have their jets incrementally more aligned to our line of sight.
             
Besides orientation, other differences also exist. \citet{Fanaroff1974} were the first to note a morphological divide in radio jets of RGs. The lower-luminosity Fanaroff-Riley type I (FRI) RGs exhibit plume-like radio lobes whereas the higher-luminosity Fanaroff-Riley type II (FRII) RGs exhibit collimated jets with terminal ``hotspots''. A division in total radio power was found to occur at $ {L_{178~MHz} = 2 \times 10^{25}~W~Hz^{-1}sr^{-1}}$ by Fanaroff and Riley but has lately been disputed \citep[e.g.,][]{Mingo2019}. Apart from total jet power, the combined effects of radiative losses, particle composition, and environmental effects on kiloparsec scales, are now being used to distinguish between the two classes \citep{Croston2018}. The circumstances that lead to the formation of an FRI or an FRII-like jet are still unknown. RL unification hypothesizes that quasars are the pole-on counterparts of FRII RGs while BL~Lac objects are the pole-on counterparts of FRI RGs \citep{Urry95}. BL Lac objects and radio-loud quasars are collectively referred to as blazars. Because of their high flux densities, they are the best sources to check the differences between the two classes with
polarization observations. 

\subsection{Differences in Polarisation}
Helical magnetic (B-) fields propagating outward with the jet plasma are predicted by several theoretical models of AGN jets \citep{Meier2001,Lyutikov2005,Hawley2015}. Magnetic fields are believed to play a critical role in bulk acceleration and jet propagation and are important to understand and differentiate between different jet formation mechanisms. Early polarization studies of AGN jets \citep[e.g.,][]{Bridle1984} have found differences between the two FR types. They found that the inferred B-field structures \citep[deduced to be perpendicular to the electric vector position angles, EVPA, for optically thin emission, and parallel to the EVPA for optically thick emission;][]{Pacholczyk70} could be categorised as (i) B-fields predominantly parallel to the jet axis, (ii) B-fields predominantly perpendicular to the jet axis, and, (iii) B-fields perpendicular to the jet axis at the centre of the jet, but parallel near one or both of its edges. 

FRII sources tend to have B-fields parallel to the jet axis whereas FRI sources typically display jets with the other two B-field configurations \citep{Willis1981,Bridle1982}. In the jets of the most powerful quasars, the inferred B-field direction after correction for Faraday rotation is normally along the jet, often following bends in the jet very closely \citep{Bridle1994}. Very long baseline interferometry (VLBI) observations of blazars have found differences in parsec scale B-field structures as well \citep{Cawthrone1993,Kharb2008}. BL~Lacs tend to have their parsec scale EVPA parallel to the jet direction whereas RL quasars show EVPA perpendicular to the jet direction \citep{Lister2005,Lister2013}. Also, BL~Lacs show systematically higher parsec-scale rotation measures (RM) than the quasars \citep{Zavala2005}. The distances to which such B-field structures persist are still unknown and are critical to understanding the interaction between jets and their environments at kiloparsec-scales where FRI jets decelerate \citep{Bicknell1994,Laing2002}.

\subsection{Kiloparsec-scale Radio Polarization Observations}
We have been carrying out polarization-sensitive kiloparsec-scale radio observations with the upgraded Giant Metrewave Radio Telescope (uGMRT) and the Karl G. Jansky Very Large Array (VLA) for a sample of blazars to study their kiloparsec-scale B-field structures. The sample chosen for this study is a subsample of the Palomar Green UV-excess photographic survey carried out in the late 1970s \citep{Green1986}. The PG UV-excess survey has no radio selection biases and is still the most extensive complete optically selected survey for unobscured AGN at low redshift. This makes it a useful sample to analyse the FR dichotomy, which was first noted in the radio-selected 3C sample. The PG sample is one of the most thoroughly researched AGN samples, \citep{Boroson1992,Kellermann1989,Miller1993}  and it has a wealth of supplemental multi-band data that include precise black hole (BH) masses \citep{Kaspi2000, Wu_2009,Shangguan2019}, accretion rates \citep{Davis2011}, and galactic properties like star-formation rates (SFR) \citep{Shi2014,Xie2021} and host galaxy morphologies \citep{Kim2017}.

The PG ``blazar'' sample \citep[see][]{Baghel2023} comprises 16 RL quasars and 8 BL~Lac objects chosen using the following selection criteria:

\vspace{-\baselineskip}
\begin{enumerate}
\item Redshift $z<0.5$
\item Projected core-to-lobe extents $\gtrsim15''$ 
\end{enumerate}

The projected lobe extents were selected to allow at least three samplings of the lobes by $\sim5''$ uGMRT Band~4 (${\sim 650~MHz}$) observations. The $\sim1''$ VLA $6~GHz$ B-array observations
 as a result probe regions of at least $\sim5~kpc$ scale at these distances. Below, we present the results from the VLA imaging of nine of the quasars and preliminary results from the uGMRT images of the BL~Lac objects.

\begin{figure}
\centering
\includegraphics[width=8.5cm,trim=150 240 150 240]{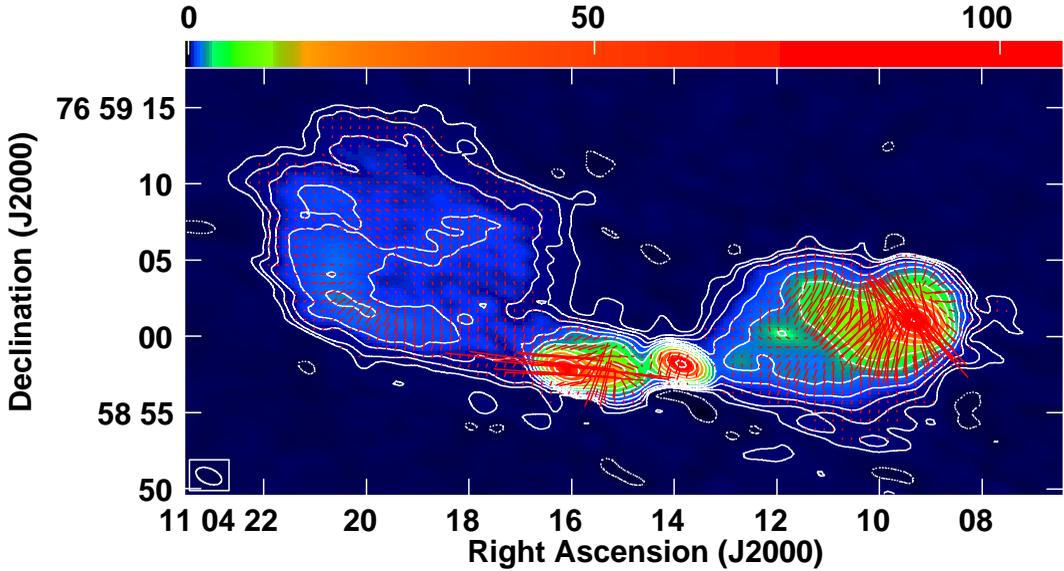}
\caption{{ VLA 6 GHz total intensity contour (and color) image of quasar PG1100+772 superimposed with red EVPA vectors. The inferred B-fields are perpendicular to the EVPA vectors assuming optically thin emission. The beam is of size $1.79'' \times 1.00''$ with a PA of $67.35^\circ$. The peak surface brightness, $I_P$ is $107.6~mJy~beam^{-1}$. The contour levels in percentage of $I_P$ are $(-0.09,~0.09,~0.18,~0.35,~0.7,~1.4,~2.8,~5.6,~11.25,~22.5,~45,~90)~Jy~beam^{-1}$. The length of the EVPA vectors is proportional to polarized intensity with $5''$ corresponding to $3.57~mJy~beam^{-1}$ \citep{Baghel2023}. }
}
\label{fig1}
\end{figure}

\section{Results} 
We find that several PG quasars display distorted (S- or X-shaped) morphologies, hybrid FR radio morphologies, or restarted AGN activity \citep{Baghel2023}. We attribute this to the PG sample's optical/UV selection criteria, which remain unbiased at radio frequencies. VLA observations find evidence of ordered B-fields on kiloparsec-scales in the PG quasars. The linear fractional polarization ranges from 1-10\% in the radio cores to 10-40\% in the jets or lobes. The radio cores typically exhibit B-fields transverse to, and jets display B-fields typically aligned with, the jet directions. Hotspots exhibit either transverse B-field signifying compression at terminal shocks or complex B-field structures. BL~Lacs and quasars do not fit neatly into the FR total power regimes at 150~MHz for the PG ``blazar'' sample. The VLA in-band (4.5 - 6.6~GHz) spectral indices of the cores are relatively flat but are steep in hotspots, consistent with polarization structures where the hotspots appear to be locations of jet bends or bow-shocks. Preliminary results from the uGMRT measurements at $\sim650~MHz$ reveal more diffuse polarised emission in blazars and indicate differences in the kiloparsec-scale B-field structures of quasars and BL~Lacs.

\section{Discussion and Conclusions}
The distorted/hybrid/restarted radio structures could be a consequence of the selection criteria of the PG sample which relies on UV-excess and a star-like nucleus. These features are strongly tied to the accretion disk state and are basically unbiased in terms of the radio properties of these sources. High sensitivity, low-frequency radio data with less restrictive selection effects than previous studies also suggest a more complicated extended source population, which includes possible hybrid radio galaxies, restarting and remnant radio galaxies \citep[e.g.,][]{Kapinska2017,Mingo2019,Jurlin2021}. In principle, this high level of diversity in radio morphology could imply that kiloparsec-scale radio emission is highly influenced by interaction with its environment. We find that the entire PG ``blazar" sample, including all the subtypes, does not cleanly fit into the FR total power regimes at 150 MHz, as well as FR radio morphologies. Several quasar lobes lack the terminal hotspot that is characteristic of FRII radio galaxies \citep[e.g.,][]{Kharb10}.

A significant portion of our sources are radio-intermediate \citep[with $10<R<250$;][]{Falcke1996b}, however we find that extensive polarization is seen in these sources  up to kiloparsec scales. Assuming that the aligned B-fields in the jets signify the dominance of a poloidal B-field component in them, and the transverse B-fields in the cores signify a dominant toroidal B-field component at the unresolved bases of the jets, then we see a transition from a toroidal field to a poloidal field along the radio jets \citep[see also III~Zw~2;][]{Silpa2021}. Since the poloidal field varies as $r^{-2}$ and the toroidal field varies as $r^{-1}~v^{-1}$, it has been hypothesized \citep[e.g.,][]{Begelman1984} that the toroidal B-field component becomes more dominant and the poloidal component decays quicker with distance from the core. In order to explain the quasar results, one must therefore invoke a mechanism that can maintain poloidal B-fields across considerable distances, such as the cosmic battery mechanism \citep{Contapoulos2009}. 

In conclusion, as higher sensitivity radio data are unveiling more complex structures in both the radio morphologies and B-fields, study of the jet-medium interactions in detail is imperative to gain a better understanding of the Fanaroff-Riley and blazar divide.

\begin{multicols}{2}
{\def\bibfont{\small}
\setlength{\bibsep}{0pt}
\bibliographystyle{astron}
\bibliography{references} }
\end{multicols}

\end{document}